\documentclass[aps,twocolumn,pra,superscriptaddress,showpacs,tightenlines]{revtex4}
\usepackage{amsmath}
\usepackage{graphicx}
\usepackage{epsfig}
\usepackage{amsfonts}

\begin{document}

\title{Correlated two-photon transport in a one-dimensional waveguide side-coupled to a nonlinear cavity}
\author{Jie-Qiao Liao}
\affiliation{Department of Physics and Institute of Theoretical
Physics, The Chinese University of Hong Kong, Shatin, Hong Kong
Special Administrative Region, People's Republic of China}
\author{C. K. Law}
\affiliation{Department of Physics and Institute of Theoretical
Physics, The Chinese University of Hong Kong, Shatin, Hong Kong
Special Administrative Region, People's Republic of China}

\begin{abstract}
${\O}$ We investigate the transport properties of two photons inside
a one-dimensional waveguide side-coupled to a single-mode nonlinear
cavity. The cavity is filled with a nonlinear Kerr medium. Based on
the Laplace transform method, we present an analytic solution for
the quantum states of the two transmitted and reflected photons,
which are initially prepared in a Lorentzian wave packet. The
solution reveals how quantum correlation between the two photons
emerge after the scattering by the nonlinear cavity. In particular,
we show that the output wave function of the two photons in position
space can be localized in relative coordinates, which is a feature
that might be interpreted as a two-photon bound state in this
waveguide-cavity system.
\end{abstract}

\maketitle

\section{\label{introduction}Introduction}

Creating quantum correlations among photons has been a subject of
major interest for studying the foundations of quantum theory as
well as applications in quantum information science. As direct
interactions between photons in free space are extremely weak,
generation of correlated photons generally requires nonlinear media.
Electromagnetically induced transparency and photon blockade are
mechanisms that have been exploited to achieve strongly interacting
photons~\cite{Schmidt1996,Deutsch1997,Harris1998,Kimble2005}.
Recently, studies of two-photon scattering from a two-level system
inside a one-dimensional (1D) waveguide have also reported various
features of photon correlation \cite{Kojima2003,Fan2007,Roy20010}.
For example, Shen and Fan~\cite{Fan2007} have discovered the
existence of two-photon bound states, and Roy \cite{Roy20010} has
indicated an interesting application of the system as a few-photon
optical diode. We also note that Shi and Sun~\cite{Shi2009} have
employed a formal scattering theory to study multi-photon transport
in a 1D waveguide.

In this paper, we investigate the correlation properties of two
photons in a 1D waveguide that is side-coupled to a nonlinear cavity
filled with a Kerr medium (Fig.~\ref{setup}). The nonlinear cavity
plays the role of a scatterer. It is worth noting that such a Kerr
nonlinearity has also been employed in coupled cavity array systems
for studying quantum phase transition
\cite{Hartmann,Greentree2006,Angelakis2007,
Hartmann2010,Yamamoto2008,Tomadin2010} and nonclassical photon
statistics~\cite{Fazio2009,Savona2010,Ferretti2010,Bamba2010}. Here
we will focus on the transport properties of two photons determined
by the long time solution of the Schr\"odinger equation, assuming
the initial photons are in wave packet forms. We will present an
analytic solution based on the Laplace transform method, which has
been applied to related photon-atom scattering problems~\cite{tsoi}.
From the two-photon transmission and reflection amplitudes, we show
how the two scattered photons can be correlated in frequency and
position variables, with the latter revealing photon bunching and
anti-bunching effects. Our solution also reveals a two-photon
resonance condition when the incident photon energies match the
cavity frequency shifted by the Kerr interaction. The behavior of
transmission and reflection near the resonance will be discussed.

\section{\label{Sec:2}Physical model}

\begin{figure}[tbp]
\includegraphics[bb=54 538 533 680, width=3.2 in]{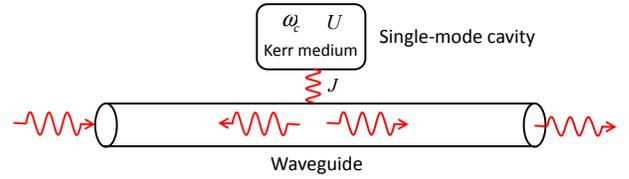}
\caption{(Color online). Schematic diagram of the physical setup. A
1D waveguide is coupled to a cavity filled with a Kerr-type
nonlinear medium. Photons injected from the left-hand side of the
waveguide are scattered by the nonlinear cavity. As a result,
photons are reflected or transmitted in the waveguide.}
\label{setup}
\end{figure}

The physical model under investigation consists of an infinitely
long 1D waveguide and a nonlinear cavity located at the origin
(Fig.~\ref{setup}).
We consider a single-mode field in the cavity, which couples to
right- and left-propagating fields of the waveguide via the side
coupling~\cite{Yariv2000,Fan2004} so that photons can tunnel between
the waveguide and the nonlinear cavity. The Hamiltonian (with
$\hbar=1$) of the system is given by
\begin{eqnarray}
\hat{H}&=&\omega _{c}\hat{a}^{\dagger
}\hat{a}+\frac{U}{2}\hat{a}^{\dagger }\hat{a}^{\dagger
}\hat{a}\hat{a}+\int_{0}^{\infty}dk\omega _{k}( \hat{r}_{k}^{\dagger
}\hat{r}_{k}+\hat{l}_{k}^{\dagger
}\hat{l}_{k})\nonumber\\
&&+J\int_{0}^{\infty}dk\left[\hat{a}^{\dagger}(
\hat{r}_{k}+\hat{l}_{k})+( \hat{r}_{k}^{\dagger
}+\hat{l}_{k}^{\dagger }) \hat{a}\right].\label{H_tot}
\end{eqnarray}
Here $\hat{a}$ and $\hat{a}^{\dagger }$ are annihilation and
creation operators associated with the cavity mode with the
resonance frequency $\omega _{c}$. The second term in
Eq.~(\ref{H_tot}) describes the Kerr nonlinear interaction with
strength $U$. The Hamiltonian of free fields propagating in the
waveguide is described by the third term, where $\hat{l}_{k}$
($\hat{l}^{\dagger}_{k}$) and $\hat{r}_{k}$
($\hat{r}^{\dagger}_{k}$) are, respectively, the annihilation
(creation) operators for left- and right-propagating waves with wave
number $k$ and frequency $\omega_{k}$. These operators satisfy the
commutation relations
\begin{eqnarray}
[\hat{l}_{k},\hat{l}^{\dagger}_{k'}]=[\hat{r}_{k},\hat{r}^{\dagger}_{k'}]=\delta(k-k'),
\hspace{0.5 cm} [\hat{l}_{k},\hat{r}^{\dagger}_{k'}]=0.
\end{eqnarray}
Finally, the last term in the Hamiltonian~(\ref{H_tot}) represents
the coupling between the cavity and the waveguide, where $J$ is the
tunneling strength.

For convenience, we introduce even- and odd-parity modes operators
of the waveguide,
\begin{eqnarray}
\hat{b}_{k}\equiv\frac{1}{\sqrt{2}}(\hat{r}_{k}+\hat{l}_{k}),\hspace{0.5
cm} \hat{c}_{k}\equiv\frac{1}{\sqrt{2}}(\hat{r}_{k}-\hat{l}_{k}),
\end{eqnarray}
so that Hamiltonian~(\ref{H_tot}) can be rewritten as
\begin{eqnarray}
\hat{H}=\hat{H}^{(o)}+\hat{H}^{(e)}
\end{eqnarray}
with
\begin{subequations}
\begin{align}
\hat{H}^{(o)}=&\int_{0}^{\infty}dk\omega _{k}\hat{c}_{k}^{\dagger
}\hat{c}_{k},\\
\hat{H}^{(e)}=&\omega _{c}\hat{a}^{\dagger
}\hat{a}+\frac{U}{2}\hat{a}^{\dagger }\hat{a}^{\dagger
}\hat{a}\hat{a}+\int_{0}^{\infty}dk\omega _{k}\hat{b}_{k}^{\dagger
}\hat{b}_{k}\nonumber\\
&+g\int_{0}^{\infty}dk(\hat{a}^{\dagger}\hat{b}_{k}+
\hat{b}_{k}^{\dagger }\hat{a}).
\end{align}
\end{subequations}
Here $g\equiv\sqrt{2}J$ is introduced. We see that the interaction
involves only even modes, and photons in the odd modes evolve freely
in the waveguide. Therefore we shall focus on the calculation of the
transport properties of the photons in even modes.

In the rotating frame with respect to $\hat{H}^{(e)}_{0}=\omega
_{c}\hat{a}^{\dagger }\hat{a}+\omega
_{c}\int_{0}^{\infty}dk\hat{b}_{k}^{\dagger }\hat{b}_{k}$,  the
Hamiltonian $\hat{H}^{(e)}$ can be simplified to
\begin{eqnarray}
\hat{H}_{I}^{(e)}&=&\frac{U}{2}\hat{a}^{\dagger }\hat{a}^{\dagger
}\hat{a}\hat{a}+\int_{0}^{\infty}dk\Delta _{k}\hat{b}_{k}^{\dagger
}\hat{b}_{k} \nonumber \\
&& +g\int_{0}^{\infty}dk(\hat{a}^{\dagger
}\hat{b}_{k}+\hat{b}_{k}^{\dagger }\hat{a}),
\end{eqnarray}
where $\Delta _{k}=\omega _{k}-\omega _{c}$ is the detuning. In this
paper the dispersion relation for the modes in the waveguide is
assumed to be linear, i.e., $\omega_k = v_g k$, and we will set the
speed of light in the waveguide as $v_g=1$.

\section{\label{Sec:3}Single-photon transport}

As a preparation for finding the solution for two-photon scattering,
we first consider the single-photon problem
~\cite{Fanpapers,Lukin-np,Law2008,Sun,Liao2009}. Note that the Kerr
nonlinearity has zero effect for single photon states. The main
purpose in this section is to present the single-photon transmission
and reflection coefficients, which will appear in the two-photon
solution later in the paper.

In the single-excitation subspace, an arbitrary state can be written
as
\begin{eqnarray}
\vert\varphi(t)\rangle =\alpha(t)\vert 1\rangle _{c}\vert
\emptyset\rangle +\int_{0}^{\infty}dk\beta _{k}(t)\vert
0\rangle_{c}\vert 1_{k}\rangle,
\end{eqnarray}
where $\vert 1\rangle _{c}\vert\emptyset\rangle$ stands for the
state with one photon in the cavity and no photon in the waveguide,
and $\vert 0\rangle_{c}\vert 1_{k}\rangle$ denotes the state with a
vacuum cavity  field and one photon in the $k$th (even) mode of the
waveguide. The time dependent variables $\alpha(t)$ and $\beta
_{k}(t)$ are the respective probability amplitudes.

By the Schr\"{o}dinger equation $i\vert\dot{\varphi}(t)
\rangle=\hat{H}_{I}^{(e)}\vert\varphi (t)\rangle$, we have
\begin{subequations}
\label{eq:eqsinpho}
\begin{align}
\dot{\alpha}(t)&=-ig\int_{0}^{\infty}dk\beta _{k}(t),\\
\dot{\beta}_{k}(t)&=-i\Delta _{k}\beta _{k}(t) -ig\alpha(t).
\end{align}
\end{subequations}
By performing the Laplace transform defined by
$\tilde{f}(s)\equiv\int_{0}^{\infty}f(t)e^{-st}dt$,
Eq.~(\ref{eq:eqsinpho}) becomes
\begin{subequations}
\begin{align}
s\tilde{\alpha}(s) -\alpha(0)&=-ig\int_{0}^{\infty}dk\tilde{
\beta}_{k}(s),\label{alpha_keq}\\
s\tilde{\beta}_{k}(s)-\beta _{k}(0) &=-i\Delta
_{k}\tilde{\beta}_{k}(s) -ig\tilde{\alpha}(s),\label{beta_keq}
\end{align}
\end{subequations}
where $\alpha(0)$ and $\beta _{k}(0)$ are the initial values of the
probability amplitudes.

Assuming that initially the cavity is in the vacuum state and an
incident single photon in the waveguide is prepared in a wave packet
with a Lorentzian spectrum, the initial condition reads
\begin{eqnarray}
\alpha \left( 0\right)  &=&0, \hspace{0.5 cm}\beta
_{k}(0)=\frac{G_{1}}{\Delta _{k}-\delta
+i\epsilon},\label{singphoini}
\end{eqnarray}
where $\delta$ and $\epsilon$ are the detuning and spectral width of
the photon, and $G_{1}=\sqrt{\epsilon/\pi}$ is a normalization
constant. The choice of $\beta_k(0)$ in Eq.~(\ref{singphoini}) has
the advantage that analytic solutions can be obtained conveniently.
In addition, by noting that $\epsilon \to 0$ corresponds to the
monochromatic limit, an incident wave packet of a general form can
be constructed by coherent superpositions of Lorentzian wave packets
of various frequencies.

After some calculations, we obtain
\begin{subequations}
\label{sinphossolu}
\begin{align}
\tilde{\alpha}(s)&=\frac{1}{s+\frac{\gamma }{2}}\frac{2\pi igG_{1}}{
\delta -i(s+\epsilon)},\\
\tilde{\beta}_{k}(s)&=\frac{G_{1}}{s+i\Delta
_{k}}\left(\frac{1}{\Delta _{k}-\delta +i\epsilon }+\frac{1}{s+
\frac{\gamma }{2}}\frac{2\gamma}{\delta-i(s+\epsilon)}\right).
\end{align}
\end{subequations}
Note that in obtaining Eq.~(\ref{sinphossolu}), we have made the
approximation: $\int_{0}^{\infty }\frac{g^{2}}{s+i\Delta
_{k}}dk\approx\int_{-\infty}^{\infty }\frac{g^{2}}{s+i\Delta_{k}}
d\Delta _{k} =\gamma/2$, where $\gamma =2\pi g^{2}$.

Taking the inverse Laplace transform of Eq.~(\ref{sinphossolu}), in
the long time limit, $\gamma t/2\rightarrow \infty $ and $\epsilon
t\rightarrow \infty$, we have
\begin{eqnarray}
\alpha(t\rightarrow\infty)&=0, \hspace{0.5 cm}\beta
_{k}(t\rightarrow\infty)=\bar{t}_{k}\beta _{k}\left( 0\right)
e^{-i\Delta _{k}t},\label{sinpholontso}
\end{eqnarray}
where
\begin{equation}
\bar{t}_{k}=\frac{\Delta _{k}-i\gamma/2}{\Delta
_{k}+i\gamma/2}.\label{defit_k}
\end{equation}
Equation~(\ref{sinpholontso}) shows that the scattering process
results in a phase shift $\theta_{k}$ for a single photon with wave
vector $k$, where the phase shift is defined by
$\exp(i\theta_{k})=\bar{t}_{k}$.

In terms of the left- and right-propagation modes, if we assume a
photon packet is incident onto the cavity from the left, then the
initial state can be written as
\begin{eqnarray}
|\varphi(0)\rangle&=&\int_{0}^{\infty}
dk\beta_{k}(0)\hat{r}^{\dag}_{k}|\emptyset\rangle\nonumber\\
&=&\frac{1}{\sqrt{2}}\int_{0}^{\infty}
dk\beta_{k}(0)(\hat{b}^{\dag}_{k}+\hat{c}^{\dag}_{k})|\emptyset\rangle.
\end{eqnarray}
In the long-time limit, the wave function becomes,
\begin{eqnarray}
|\varphi(t\rightarrow\infty)\rangle = \int_{0}^{\infty}
dk\beta_{k}(0)e^{-i\Delta_{k}t}(t_{k}\hat{r}^{\dag}_{k}
+r_{k}\hat{l}^{\dag}_{k})| \emptyset\rangle,
\end{eqnarray}
where the transmission and reflection amplitudes are defined as
\begin{eqnarray}
t_{k}=\frac{\Delta_{k}}{\Delta_{k}+i\gamma/2}, \hspace{0.5
cm}r_{k}=\frac{-i\gamma/2}{\Delta_{k}+i\gamma/2}.\label{deftkrk}
\end{eqnarray}
A similar result has been obtained for the case that a single photon
is scattered by a two-level system in a 1D waveguide
~\cite{Fanpapers}, namely, the transmission amplitude $t_k$ is zero
at the exact resonance. This effect was also reported in Ref.
\cite{Yariv2000} for side coupling with a classical field.

\section{\label{Sec:4}Correlated two-photon transport}

\subsection{Equations of motion and solution}

We now turn to the two-photon scattering problem. Since the total
excitation number operator of the system is a conserved quantity, we
can restrict the calculation to the two-excitation subspace. An
arbitrary state in this subspace has the form:
\begin{eqnarray}
\vert\Phi(t)\rangle&=&A(t)\vert 2\rangle_{c}\vert\emptyset\rangle
+\int_{0}^{\infty}dkB_{k}(t)\vert 1\rangle_{c}\vert 1_{k}\rangle\nonumber\\
&&+\int_{0}^{\infty}dp\int_{0}^{p}dqC_{p,q}(t)\vert
0\rangle_{c}\vert 1_{p},1_{q}\rangle,
\end{eqnarray}
where $\vert 2\rangle_{c}\vert\emptyset\rangle$ is the state of two
photons in the nonlinear cavity and no photon in the waveguide, and
$\vert 1\rangle_{c}\vert 1_{k}\rangle$ is the state with one photon
in the cavity and one photon with wave number $k$ in the waveguide.
The last term represents the state with no photon in the cavity and
two photons with wave numbers $p$ and $q$ in the waveguide. $A(t)$,
$B_{k}(t)$, and $C_{p,q}(t)$ denote the respective probability
amplitudes.

By the Schr\"{o}dinger equation, the probability amplitudes are
governed by:
\begin{subequations}
\label{eqfortwophoto}
\begin{align}
& \dot{A}\left( t\right) =-iUA\left( t\right)
-i\sqrt{2}g\int_{0}^{\infty}dkB_{k}\left(
t\right),\\
& \dot{B}_{k}(t) =-i\Delta _{k}B_{k}(t) -i\sqrt{2}
gA\left( t\right) -ig\int_{0}^{\infty}dpC_{p,k}(t),\\
& \dot{C}_{p,q}\left( t\right) =-i(\Delta_{p}+\Delta_{q}) C_{p,q}(t)
-ig(B_{p}\left( t\right) +B_{q}\left( t\right)).
\end{align}
\end{subequations}

We assume that the two injected photons are initially prepared in a
Lorentzian wave packet. The initial condition of the system reads,
\begin{subequations}
\label{inicondfortwophoto}
\begin{align}
A\left( 0\right)=&0,\hspace{0.5 cm}
B_{k}\left( 0\right)=0,\\
C_{p,q}\left( 0\right) =&G_{2}\left( \frac{1}{\Delta _{p}-\delta
_{1}+i\epsilon } \frac{1}{\Delta _{q}-\delta _{2}+i\epsilon
}\right.\nonumber\\
&\left.+\frac{1}{\Delta _{q}-\delta _{1}+i\epsilon }\frac{1}{\Delta
_{p}-\delta _{2}+i\epsilon }\right),\label{cpqinitcon}
\end{align}
\end{subequations}
with the normalization constant
\begin{eqnarray}
G_{2}=\frac{\epsilon }{\sqrt{2}\pi}\left(1+\frac{4\epsilon
^{2}}{(\delta _{1}-\delta _{2})^{2}+4\epsilon^{2}}\right)^{-1/2}.
\end{eqnarray}
Here $\delta_j$ and $\epsilon_j$ ($j=1,2$) are parameters defining
the detunings and spectral widths of the two photons. Note that
$C_{p,q}$ has been symmetrized in Eq.~(\ref{cpqinitcon}) because of
the bosonic character of photons.

We are interested in the asymptotic solution of $C_{p,q}(t)$ in the
long time limit. After a lengthy calculation (see
Appendix~\ref{appa}), we obtain for $ t \gg \gamma ^{-1}$ and
$\epsilon^{-1}$,
\begin{eqnarray}
C_{p,q}(t)=\left(\bar{t}_{p}\bar{t}_{q}C_{p,q}(0)+B_{p,q}\right)
e^{-i(\Delta_{p}+\Delta_{q})t},\label{definiC_pq}
\end{eqnarray}
where $\bar{t}_{p}$ and $\bar{t}_{q}$ are defined in
Eq.~(\ref{defit_k}). The expression of $B_{p,q}$ is given by
\begin{eqnarray}
B_{p,q} &=&\frac{-2UG_{2}\gamma ^{2} }{\left( \Delta
_{p}+i\frac{\gamma }{2}\right) \left( \Delta _{q}+i\frac{\gamma }{2}
\right)( \Delta _{p}+\Delta
_{q}-U+i\gamma)}\nonumber\\
&&\times\frac{1}{(\Delta _{p}+\Delta _{q}-\delta _{1}-\delta
_{2}+2i\epsilon) }\nonumber\\
&& \times \left[\frac{1}{\left( \Delta _{p}+\Delta _{q}-\delta
_{1}+i\epsilon +i \frac{\gamma }{2}\right)
}\right.\nonumber\\
&&\left.+\frac{1}{\left( \Delta _{p}+\Delta _{q}-\delta
_{2}+i\epsilon +i\frac{\gamma }{2}\right) }\right] .\label{exprBpq}
\end{eqnarray}
From Eqs.~(\ref{definiC_pq}) and~(\ref{exprBpq}), we notice that the
term $B_{p,q}$ is a non-factorizable function of $p$ and $q$,
implying a correlation between the two output photons. $B_{p,q}$ has
a numerator proportional to the strength of the Kerr nonlinearity
$U$ in the cavity. In the case $U=0$, Eq.~(\ref{definiC_pq}) reduces
to a simple expression
$C_{p,q}(\infty)=\bar{t}_{p}\bar{t}_{q}C_{p,q}(0)\exp[-i(\Delta_{p}+\Delta_{q})t]$,
describing two independent scattered photons.

\subsection{Two-photon correlation in frequency variables}

Let us express the results in terms of the left- and
right-propagating modes. Assuming the two photons are injected from
the left-hand side of the waveguide, then the initial wave function
can be written as
\begin{eqnarray}
\vert \psi (0)\rangle=\int_{0}^{\infty} \int_{0}^{\infty}
dpdqC_{p,q}\left( 0\right) \hat{r}_{p}^{\dagger
}\hat{r}_{q}^{\dagger }\left\vert \emptyset \right\rangle
\end{eqnarray}
According to Eq. (3) and the solution (\ref{definiC_pq}), we obtain
the long-time wave function, up to an overall phase factor
$\exp[-i(\Delta_{p}+\Delta_{q})t]$, as
\begin{eqnarray}
\left\vert \psi \left( t\rightarrow \infty \right) \right\rangle &=&
\int_{0}^{\infty} \int_{0}^{\infty}dpdq (
C_{p,q}^{rr}\hat{r}_{p}^{\dagger }\hat{r}_{q}^{\dagger } +
C_{p,q}^{ll}\hat{l}
_{p}^{\dagger }\hat{l}_{q}^{\dagger } ) \vert \emptyset \rangle \nonumber  \\
&& +\int_{0}^{\infty} \int_{0}^{\infty}dpdq
(C_{p,q}^{rl}\hat{r}_{p}^{\dagger }\hat{l}_{q}^{\dagger }
+C_{p,q}^{lr}\hat{l} _{p}^{\dagger }\hat{r}_{q}^{\dagger
})\vert \emptyset\rangle,\nonumber\\
\label{longtsolu}
\end{eqnarray}
where
\begin{subequations}
\label{cpqintwomodes}
\begin{align}
\label{cpqintwomodes_a}C_{p,q}^{rr} &=t_{p}t_{q}C_{p,q}\left( 0\right) +\frac{1}{4}B_{p,q},\\
C_{p,q}^{ll} &=r_{p}r_{q}C_{p,q}\left( 0\right) +\frac{1}{4}B_{p,q},\\
C_{p,q}^{rl} &=t_{p}r_{q}C_{p,q}\left( 0\right) +\frac{1}{4}B_{p,q},\\
C_{p,q}^{lr} &=t_{q}r_{p}C_{p,q}\left( 0\right) +\frac{1}{4}B_{p,q}.
\end{align}
\end{subequations}
Here $C_{p,q}^{rr}$ and $C_{p,q}^{ll}$ are, respectively, the
two-photon transmission and two-photon reflection amplitudes, which
correspond to the processes in which two photons with wave numbers
$p$ and $q$ are transmitted into the right-propagation mode or
reflected into the left-propagation mode. In addition,
$C_{p,q}^{rl}$ ($C_{p,q}^{lr}$) relates to the process where the
photon with wave number $p$ ($q$) is transmitted into the
right-propagation mode and the photon with wave number $q$ ($p$) is
reflected into the left-propagation mode.

\begin{figure}[tbp]
\center
\includegraphics[bb=49 190 552 600, width=3 in]{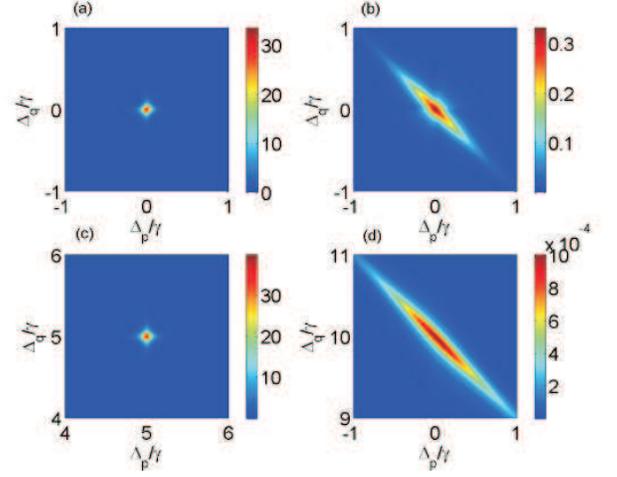}
\caption{(Color online). (a) and (b) are plots of $|\gamma
C_{p,q}^{ll}|^{2}$ and $|\gamma C_{p,q}^{rr}|^{2}$, respectively,
when $\delta_{1}=\delta_{2}=0$. (c) and (d) are plots of $|\gamma
C_{p,q}^{rr}|^{2}$ and  $|\gamma C_{p,q}^{ll}|^{2}$, respectively,
when $\delta_{1}=\delta_{2}=U/2$. Other parameters are set as
$U/\gamma=10$ and $\epsilon/\gamma=0.05$.}
\label{correlawavevectspac}
\end{figure}

We point out two interesting situations revealing the strong
correlation of output photons in the frequency domain. The first
situation is achieved by injecting two identical photons with
$\delta_1=\delta_2=0$ and a narrow spectral width $\epsilon \ll
\gamma$. This corresponds to the case when the peak frequency of the
photons coincides with the resonant cavity. In this case the two
photons are mainly reflected and uncorrelated
[Fig.~\ref{correlawavevectspac}(a)], but if they are transmitted,
they are strongly correlated [Fig.~\ref{correlawavevectspac}(b)].
This can be seen by the fact that $t_p=t_q=0$ at zero detuning, and
hence the transmission of both photons is dominated by the $B_{p,q}$
term. In other words, the two-photon transmission near
$\delta_1=\delta_2=0$ is almost entirely due to the nonlinearity in
the cavity. Such a pair of transmitted photons is frequency
correlated with the two-photon transmission probability concentrated
along the line $\Delta_p + \Delta_q =0$
[Fig.~\ref{correlawavevectspac}(b)]. The uncertainty in the
frequencies of individual transmitted photons is of the order of
$\gamma$, whereas the uncertainty in the sum of the frequencies of
both photons is of the order of $\epsilon$. The smaller $\epsilon$,
the narrower is the distribution.

The second situation of interest is two-photon resonance occurring
when the sum of energies of the two incident photons equals to the
energy of a cavity containing two photons, i.e.,
$\delta_{1}+\delta_{2}=U$. In this case the photons can jointly
enter the cavity. We show in Figs.~\ref{correlawavevectspac}(c) and
~\ref{correlawavevectspac}(d) an example with $\delta_1=\delta_2=U/2
\gg \gamma$, where the frequency correlation appears more
effectively in the reflected amplitude $C_{p,q}^{rr}$, since $r_j
\approx 0$ ($j=p,q$). This is shown in the narrow distribution in
Fig.~\ref{correlawavevectspac}(d). The transmission part
[Fig.~\ref{correlawavevectspac}(c)], although they carry most of the
probabilities, are almost uncorrelated.

\subsection{Two-photon correlation in position variables}

\begin{figure}[tbp]
\center
\includegraphics[bb=14 69 419 592, width=2.45 in]{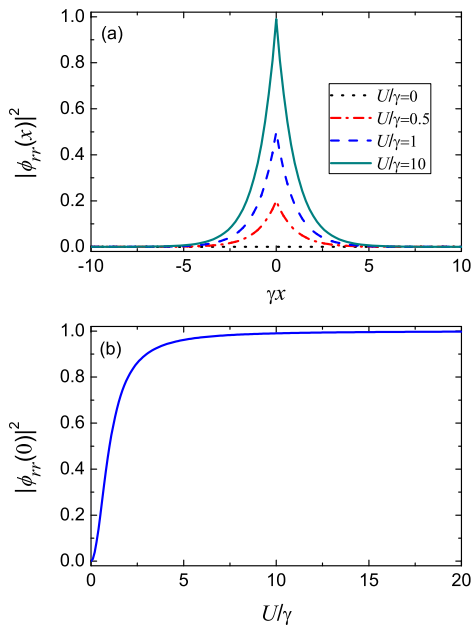}
\caption{(Color online). Spatial features of two-photon transmission
at $\delta_1=\delta_2=0$. (a) $|\phi_{rr}(x)|^{2}$ as a function of
scaled relative coordinate $\gamma x$, for various values of the
scaled Kerr parameter $U/\gamma$. (b) $|\phi_{rr}(0)|^{2}$ as a
function of the scaled Kerr parameter $U/\gamma$.}
\label{relawavefunctrans}
\end{figure}
We now discuss the spatial features of the output photons. For
simplicity, but without loss of generality, we consider the
monochromatic limit $\epsilon\rightarrow0$ of incident photons. The
two-photon transmission amplitude projected in position space reads
(see Appendix~\ref{appb})
\begin{eqnarray}
\langle x_{1},x_{2}|\psi_{rr}\rangle\approx-16\pi^{2}
\mathcal{M}G_{2}e^{iE(
x_{c}-t)}\theta(t-x_{c})\phi_{rr}(x),\label{eq25}
\end{eqnarray}
with
\begin{eqnarray}
\phi_{rr}(x)&=&t_{\delta _{1}}t_{\delta _{2}}\cos( \delta
x)-\frac{U}{E-U+i\gamma
}\nonumber\\
&&\times\frac{\gamma ^{2}}{(E +i\gamma) ^{2}-4\delta ^{2}}
e^{\frac{(iE-\gamma)}{2}\vert x\vert}. \label{phirrx}
\end{eqnarray}
Here we have defined $x_{c}=(x_{1}+x_{2})/2$ and $x=x_{1}-x_{2}$ for
the center-of-mass and relative coordinates respectively, and
$E=\delta_{1}+\delta_{2}$ and $\delta=(\delta_{1}-\delta_{2})/2$. We
note that Eq.~(\ref{eq25}) is a product of the center-of-mass wave
function and the relative wave function $\phi_{rr}(x)$, with
$\exp[iE(x_{c}-t)]\theta(t-x_{c})$ describing the center-of-mass
motion of the two transmitted photons. The second term of
$\phi_{rr}(x)$ is a function localized around $x=0$ with a width
$\gamma^{-1}$. We remark that a similar feature was reported in
Ref.~\cite{Fan2007} in a photon-atom scattering problem, where the
exponential decaying function is connected to the existence of
photon bound states.

To reveal spatial correlations, we take $\delta_{1}=\delta_{2}=0$ so
that the first term of $\phi_{rr}(x)$ can be suppressed. This is
shown in Fig.~\ref{relawavefunctrans}(a) for various values of $U$.
Note that $|\phi_{rr}(x)|^2$ is proportional to the joint
probability of photons with a separation $x$, therefore the decaying
feature corresponds to photon bunching. In particular, the joint
probability of having both transmitted photons at the same position
increases with increasing $U$, but it saturates when $U \gg \gamma$
[Fig.~\ref{relawavefunctrans}(b)].
\begin{figure}[tbp]
\center
\includegraphics[bb=9 75 416 590, width=2.5 in]{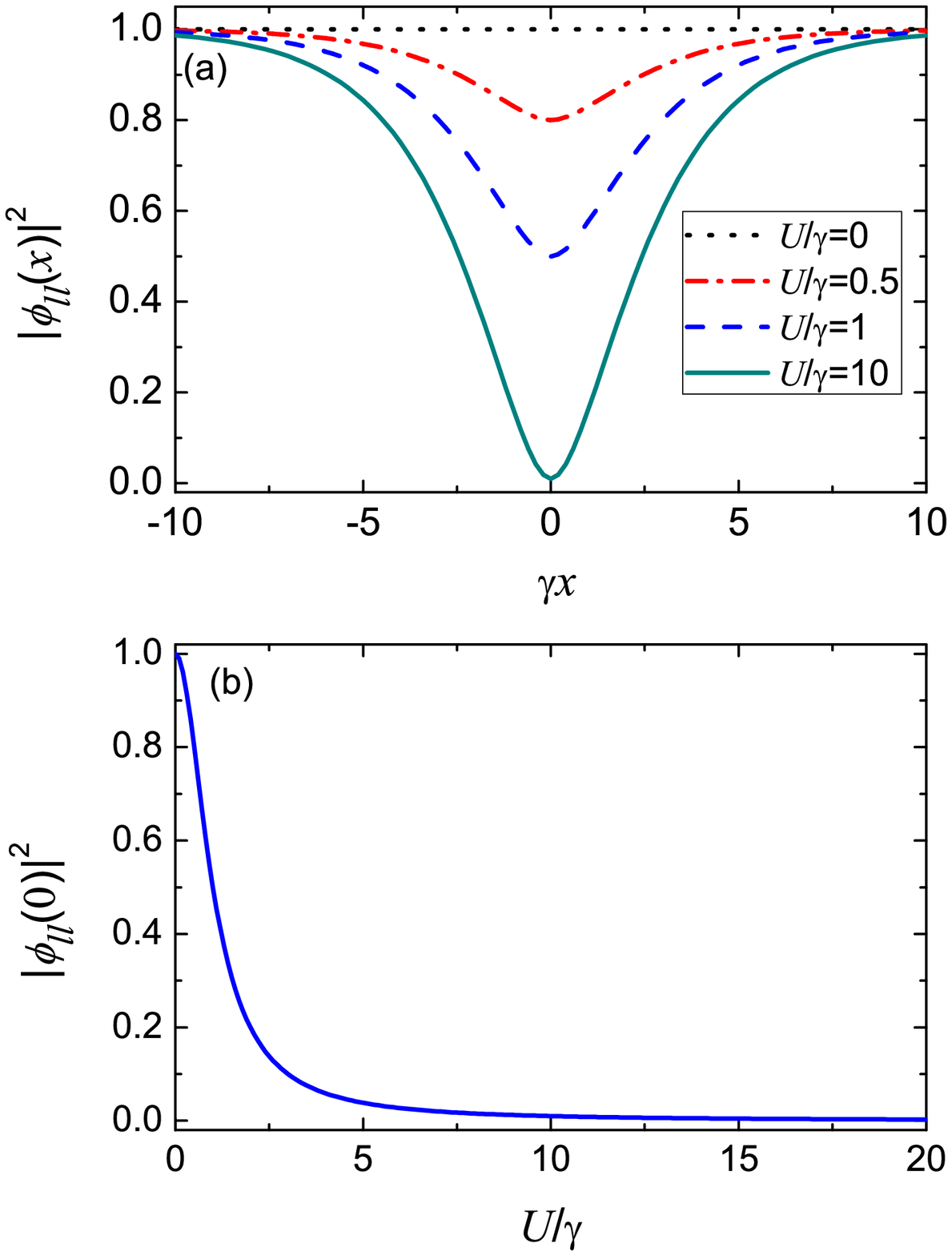}
\caption{(Color online). Spatial features of two-photon reflection
at $\delta_1=\delta_2=0$. (a) $|\phi_{ll}(x)|^{2}$ as a function of
scaled relative coordinate $\gamma x$, for various values of the
scaled Kerr parameter $U/\gamma$. (b) $|\phi_{ll}(0)|^{2}$ as a
function of the scaled Kerr parameter $U/\gamma$.}
\label{relawavefuncrefl}
\end{figure}
For the two-photon reflection amplitude in position space, we carry
out a similar calculation and obtain,
\begin{equation}
\langle x_{1},x_{2}|\psi_{ll}\rangle\approx-16\pi^{2}
\mathcal{N}G_{2}e^{-iE(x_{c}+t)}\theta(t+x_{c})\phi_{ll}(x),
\end{equation}
with
\begin{eqnarray}
\phi_{ll}(x)&=&r_{\delta _{1}}r_{\delta _{2}}\cos(
\delta x)-\frac{U}{E-U+i\gamma }\nonumber\\
&&\times\frac{\gamma^{2}}{(E+i\gamma) ^{2}-4\delta
^{2}}e^{\frac{(iE-\gamma)}{2}\vert x\vert}.\label{phillx}
\end{eqnarray}
At $\delta_{1}=\delta_{2}= \delta = 0$, the second term causes a dip
in $|\phi_{ll}(x)|^{2}$ at $x=0$ [Fig.~\ref{relawavefuncrefl}(a)],
which is a signature of photon antibunching as the reflected photons
repel each other. As $U$ increases, the joint probability of having
both reflected photons at the same position decreases
[Fig.~\ref{relawavefuncrefl}(b)], which is in contrast to the
transmitted part.

\begin{figure}[tbp]
\center
\includegraphics[bb=48 166 566 620, width=3 in]{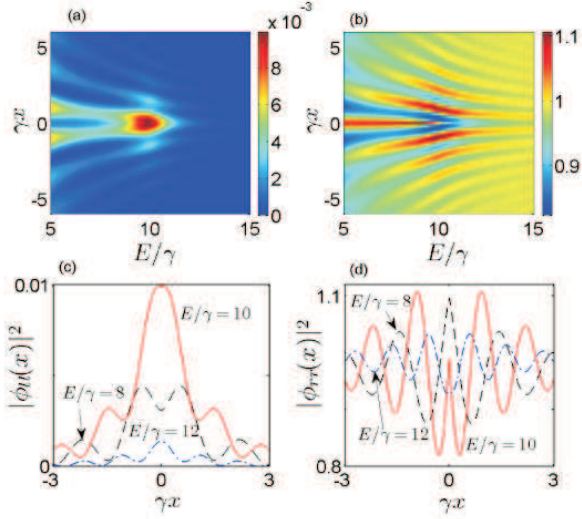}
\caption{(Color online). Dependence of spatial features on
$E=\delta_1+\delta_2$. (a) $|\phi_{ll}(x)|^{2}$ and (b)
$|\phi_{rr}(x)|^{2}$. Examples at particular values of $E$ are shown
in (c) and (d). In these figures, we use $U=10 \gamma$ and
$\delta=0$.} \label{twophotonresonant}
\end{figure}

Finally we describe the effects of two-photon resonance around
$\delta_{1}+\delta_{2}=U$ discussed in the previous section. For
simplicity we again consider the case $\delta_{1}=\delta_{2}$ here.
In Fig.~\ref{twophotonresonant}, we illustrate the dependence of the
relative two-photon wave function on $E=\delta_{1}+\delta_{2}$. The
effect of two-photon resonance is most apparent in
Fig.~\ref{twophotonresonant}(a), where the reflected two-photon wave
function is strongly localized around $x=0$ when $E=U$. Away from
the resonance, the reflected two-photon wave function exhibits an
oscillatory pattern in $x$ [Fig.~\ref{twophotonresonant}(c)], which
is controlled by the two-photon detuning $E-U$. We also plot the
transmitted two-photon wave function in
Figs.~\ref{twophotonresonant}(b) and~\ref{twophotonresonant}(d), in
which similar oscillatory patterns are observed.

\section{\label{Sec:5}Conclusions}

In conclusion, we have presented an analytic solution of two-photon
scattering inside a one-dimensional waveguide that is side-coupled
to a Kerr-type nonlinear cavity. The system provides a scheme to
realize correlated two-photon transport. The Kerr nonlinearity is
found to correlate photons in frequency variables such that
$\Delta_p + \Delta_q$ is a constant, which is a constraint because
of the energy conservation. In position space, we have shown that
the Kerr nonlinearity can cause the two photons `stick' together
with an average separation distance of the order of $v_g
\gamma^{-1}$. We may interpret the result as a two-photon bound
state, because of the exponential decaying shape of relative wave
function. However, because of the interference with single photon
processes described by the first term in Eq. (25), features of
photon correlation may only be observed efficiently in certain
directions. Finally, we note that recent studies of a related topic
have considered using a single atom as a scatterer
\cite{Fan2007,sunnew}. However, in view of recent progress in
achieving a giant Kerr nonlinearity
\cite{Schmidt1996,Deutsch1997,Harris1998}, our work suggests that a
nonlinear cavity may be an alternative way of regarding the
correlated two-photon transport problem.

\begin{acknowledgments}
This work is supported by the Research Grants Council of Hong Kong,
Special Administrative Region of China (Project No.~401408).
\end{acknowledgments}

\appendix
\begin{widetext}
\section{\label{appa}Solution of Eq.~(\ref{eqfortwophoto}) by the Laplace
transform method}

In this appendix, we give a detailed derivation for the solution of
Eq.~(\ref{eqfortwophoto}) which governs the transport of two photons
in a waveguide. We use the Laplace transform method to solve these
equations. Under the initial condition~(\ref{inicondfortwophoto}),
Eq.~(\ref{eqfortwophoto}) becomes
\begin{subequations}
\begin{align}
(s+iU)\tilde{A}(s)&= -i\sqrt{2}g\int_{0}^{\infty}dk
\tilde{B}_{k}(s),\label{aeq}\\
(s+i\Delta _{k})\tilde{B}_{k}(s) &=-i \sqrt{2}g\tilde{A}(s)
-ig\int_{0}^{\infty}dp\tilde{C}_{p,k}(s),\label{beq}\\
[s+i(\Delta_{p}+\Delta_{q})]\tilde{C}_{p,q}(s)
&=C_{p,q}(0)-ig(\tilde{B}_{p}(s)+\tilde{B}_{q}(s)).\label{ceq}
\end{align}
\end{subequations}
Substituting Eqs.~(\ref{aeq}) and (\ref{ceq}) into Eq.~(\ref{beq}),
and making use of the initial condition, we obtain the equation for
the variable $\tilde{B}_{k}\left( s\right)$ as
\begin{eqnarray}
[\Delta_{k}-i(s+\gamma/2)] \tilde{B} _{k}\left( s\right)
&=&\int_{-\infty}^{\infty}\left(
\frac{2g^{2}}{U-is}+\frac{g^{2}}{\Delta _{p}+\Delta _{k}-is}\right)
\tilde{B}_{p}\left( s\right) d\Delta _{p}
\nonumber\\
&&+2\pi gG_{2}\left( \frac{1}{\Delta _{k}+\delta _{1}-i\left(
s+\epsilon \right) }\frac{1}{\Delta _{k}-\delta _{2}+i\epsilon
}+\frac{1}{\Delta _{k}+\delta _{2}-i\left( s+\epsilon \right)
}\frac{1}{\Delta _{k}-\delta _{1}+i\epsilon } \right).\label{eqa5}
\end{eqnarray}
where we have made the approximation
$\int_{0}^{\infty}dp\frac{g^{2}}{s+i(\Delta_{p}+\Delta_{k})}
\approx\gamma/2$.

The solution of $\tilde{B}_{k}(s)$ Eq.~(\ref{eqa5}), by inspection,
takes the form:
\begin{eqnarray}
\tilde{B}_{k}(s)&=&\frac{2\pi gG_{2}}{\Delta
_{k}-i(s+\gamma/2)}\left(\frac{1}{\Delta _{k}+\delta _{1}-i(
s+\epsilon) }\frac{1}{\Delta _{k}-\delta _{2}+i\epsilon
}+\frac{1}{\Delta _{k}+\delta _{2}-i(s+\epsilon) }\frac{1}{\Delta
_{k}-\delta _{1}+i\epsilon
}\right)(1+\tilde{F}_{k}(s)),\label{bkandFk}
\end{eqnarray}
with
\begin{eqnarray}
\tilde{F}_{k}(s)&=&-i\gamma\left( \frac{1}{\Delta _{k}+\delta
_{1}-i(s+\epsilon) }\frac{1}{\Delta _{k}-\delta _{2}+i\epsilon
}+\frac{1}{\Delta _{k}+\delta _{2}-i(s+\epsilon) }\frac{1}{\Delta
_{k}-\delta _{1}+i\epsilon }\right)^{-1}\nonumber\\
&&\times\left[ \left( \frac{2}{U-is-i\gamma }+\frac{1}{\delta
_{2}+\Delta _{k}-i\left( s+\epsilon \right) }\right)\frac{1}{\delta
_{2}-i\left( s+\epsilon +\frac{\gamma }{2}\right) }\frac{1}{\delta
_{1}+\delta _{2}-i\left(
s+2\epsilon \right) }\right.\nonumber\\
&&\left.+\left( \frac{2}{U-is-i\gamma }+\frac{1}{\delta _{1}+\Delta
_{k}-i\left( s+\epsilon \right) }\right)\frac{1}{\delta _{1}-i\left(
s+\epsilon +\frac{\gamma }{2}\right) }\frac{1}{\delta _{1}+\delta
_{2}-i\left( s+2\epsilon \right) } \right].
\end{eqnarray}
Then from Eq.~(\ref{ceq}) we obtain the following expression
\begin{eqnarray}
\tilde{C}_{p,q}(s) &=&\frac{ G_{2}}{s+i(\Delta_{p}+\Delta
_{q})}\left[\frac{i\gamma}{s+\frac{\gamma }{2}+i\Delta _{p}}\left(
\frac{1}{s+\epsilon +i(\Delta_{p}+\delta_{1})}\frac{1}{\Delta
_{p}-\delta _{2}+i\epsilon }+\frac{1}{s+\epsilon
+i(\Delta_{p}+\delta_{2})}\frac{1}{\Delta _{p}-\delta _{1}+i\epsilon
}\right)
\right.\nonumber\\
&&\left.+\frac{i\gamma}{s+\frac{\gamma}{2}+i\Delta _{q}}\left(
\frac{1}{s+\epsilon +i(\Delta_{q}+\delta_{1})}\frac{1}{\Delta
_{q}-\delta _{2}+i\epsilon }+\frac{1 }{s+\epsilon +i(\Delta
_{q}+\delta _{2})}\frac{1}{\Delta
_{q}-\delta _{1}+i\epsilon }\right)\right.\nonumber\\
&&\left.-\frac{2\gamma ^{2}}{s+\gamma +iU}\frac{1}{s+2\epsilon
+i\left( \delta _{1}+\delta _{2}\right) } \left( \frac{1}{s+\epsilon
+\frac{\gamma }{2}+i\delta _{1}}+\frac{1}{ s+\epsilon +\frac{\gamma
}{2}+i\delta _{2}}\right) \left( \frac{1}{s+\frac{ \gamma
}{2}+i\Delta _{p}}+\frac{1}{s+\frac{\gamma }{2}+i\Delta _{q}}\right)
\right.\nonumber\\
&&\left.-\frac{\gamma ^{2}}{ s+2\epsilon +i(\delta _{1}+\delta _{2})
}\frac{1}{s+\epsilon + \frac{\gamma }{2}+i\delta _{1}}\left(
\frac{1}{s+\epsilon +i(\delta _{1}+\Delta_{p})
}\frac{1}{s+\frac{\gamma }{2}+i\Delta _{p}}+\frac{1}{ s+\epsilon
+i(\delta _{1}+\Delta _{q})}\frac{1}{s+\frac{\gamma
}{2}+i\Delta _{q}}\right)\right.\nonumber\\
&&\left.-\frac{\gamma ^{2}}{ s+2\epsilon +i(\delta_{1}+\delta _{2})
}\frac{1}{s+\epsilon + \frac{\gamma }{2}+i\delta _{2}}\left(
\frac{1}{s+\epsilon +i(\delta _{2}+\Delta _{p})
}\frac{1}{s+\frac{\gamma }{2}+i\Delta _{p}}+\frac{1}{ s+\epsilon
+i(\delta _{2}+\Delta _{q})}\frac{1}{s+\frac{\gamma
}{2}+i\Delta _{q}}\right)\right.\nonumber\\
&&\left.+\left( \frac{1}{\Delta _{p}-\delta _{1}+i\epsilon
}\frac{1}{\Delta _{q}-\delta _{2}+i\epsilon }+ \frac{1}{\Delta
_{q}-\delta _{1}+i\epsilon }\frac{1}{\Delta _{p}-\delta
_{2}+i\epsilon }\right)\right].
\end{eqnarray}
Until now, we have obtained the expression for $\tilde{C}_{p,q}(s)$.
Then we can get the expression for the probability amplitude
$C_{p,q}(t)$ by performing the inverse Laplace transform of
$\tilde{C}_{p,q}(s)$. In particular, since we are interested in the
output state of the two photons, here we present only the long-time
solution of $C_{p,q}(t\rightarrow\infty)$ as
\begin{eqnarray}
C_{p,q}(t\rightarrow\infty)=(\bar{t}_{p}\bar{t}_{q}C_{p,q}(0)+B_{p,q})e^{-i
(\Delta_{p}+\Delta_{q})t},
\end{eqnarray}
where $\bar{t}_{p}$ and $\bar{t}_{q}$ have been defined in
Eq.~(\ref{defit_k}), and the expression for the correlation term is
\begin{eqnarray}
B_{p,q} &=&\frac{-2UG_{2}\gamma ^{2} }{\left( \Delta
_{p}+i\frac{\gamma }{2}\right) \left( \Delta _{q}+i\frac{\gamma }{2}
\right)( \Delta _{p}+\Delta _{q}-U+i\gamma)}\frac{1}{(\Delta
_{p}+\Delta _{q}-\delta _{1}-\delta
_{2}+2i\epsilon) }\nonumber\\
&& \times \left[\frac{1}{\left( \Delta _{p}+\Delta _{q}-\delta
_{1}+i\epsilon +i \frac{\gamma }{2}\right) }+\frac{1}{\left( \Delta
_{p}+\Delta _{q}-\delta _{2}+i\epsilon +i\frac{\gamma }{2}\right)
}\right].\label{Bpqderivation}
\end{eqnarray}

\section{\label{appb}Derivation of two-photon output state in position space}

In this appendix, we derive the wave function of the two-photon
output state~(\ref{longtsolu}) in position space. For the two-photon
transmission process, the corresponding wave function in position
space can be written as
\begin{eqnarray}
\langle x_{1},x_{2}|\psi_{rr}\rangle &=&\int_{0}^{\infty}
\int_{0}^{\infty}dpdqC_{p,q}^{rr}\langle
x_{1},x_{2}|\hat{r}_{p}^{\dagger }\hat{r}_{q}^{\dagger }\vert
\emptyset
\rangle\nonumber\\
&\approx&\mathcal{M}\int_{-\infty }^{\infty }\int_{-\infty }^{\infty
}(t_{p}t_{q}C_{p,q}(0)+B_{p,q}/4) e^{-i(\Delta_{p}+\Delta _{q})
t}e^{i\Delta _{p}x_{1}}e^{i\Delta _{q}x_{2}}d\Delta _{p}d\Delta
_{q}+x_{1}\leftrightarrow x_{2}.\label{appeb1}
\end{eqnarray}
In Eq.~(\ref{appeb1}), symmetrization of the two photons has been
taken into account by introducing $\langle
x_{1},x_{2}|\hat{r}_{p}^{\dagger }\hat{r}_{q}^{\dagger }\vert
\emptyset \rangle=\mathcal{M}(e^{i\Delta _{p}x_{1}}e^{i\Delta
_{q}x_{2}}+e^{i\Delta _{p}x_{2}}e^{i\Delta _{q}x_{1}})$. According
to the initial condition given in Eq.~(\ref{cpqinitcon}), we can get
the expression for the independent transport part as
\begin{eqnarray}
&&\mathcal{M}\int_{-\infty }^{\infty }\int_{-\infty }^{\infty
}t_{p}t_{q}C_{p,q}\left( 0\right) e^{-i\left( \Delta _{p}+\Delta
_{q}\right) t}e^{i\Delta _{p}x_{1}}e^{i\Delta _{q}x_{2}}d\Delta
_{p}d\Delta _{q}\nonumber\\
&&=-8\pi^{2}\mathcal{M}G_{2}t_{\delta _{1}-i\epsilon}t_{\delta
_{2}-i\epsilon}e^{( iE+2\epsilon)(x_{c}-t)}\cos(\delta
x)\theta(t-x_{c}),\label{Foffreeterm}
\end{eqnarray}
where we introduce the center-of-mass coordinator $x_{c}
=(x_{1}+x_{2})/2$, the relative coordinator $x=x_{1}-x_{2}$, the
total momentum $E=\delta _{1}+\delta _{2}$, and the relative
momentum $\delta =(\delta _{1}-\delta _{2})/2$. $\theta(x)$ is the
Heaviside step function and $t_{\delta _{1}-i\epsilon}$ is defined
in Eq.~(\ref{deftkrk}). Note that here we have taken the
approximation $\exp[\gamma(x_{1}-t)/2]\rightarrow 0$ under the
assumption of $\gamma/2\gg\epsilon$.

According to Eq.~(\ref{Bpqderivation}), the Fourier transform of the
correlation part $B_{pq}$ can be written as
\begin{eqnarray}
\mathcal{M}\int_{-\infty }^{\infty }\int_{-\infty }^{\infty
}\frac{B_{p,q}}{4}e^{-i\left( \Delta _{p}+\Delta _{q}\right)
t}e^{i\Delta _{p}x_{1}}e^{i\Delta _{q}x_{2}}d\Delta _{p}d\Delta _{q}
=A_{1}+A_{2},
\end{eqnarray}
with
\begin{eqnarray}
A_{l}&=&-\frac{1}{2}\mathcal{M}UG_{2}\gamma ^{2}\int_{-\infty
}^{\infty }\int_{-\infty}^{\infty }\frac{1}{( \Delta _{q}+i\gamma/2)
}\frac{1}{(\Delta_{p}+i\gamma/2) }\frac{1}{\left( \Delta _{p}+\Delta
_{q}-U+i\gamma \right) }\frac{1}{\left( \Delta _{p}+\Delta
_{q}-\delta _{1}-\delta
_{2}+2i\epsilon \right) }\nonumber\\
&&\times\frac{1}{(\Delta _{p}+\Delta _{q}-\delta _{l}+i\epsilon
+i\gamma/2)} e^{i\Delta _{p}\left( x_{1}-t\right) }d\Delta
_{p}e^{i\Delta _{q}\left( x_{2}-t\right) }d\Delta _{q},\nonumber\\
\label{FTofBpq}
\end{eqnarray}
for $l=1,2$. The Fourier transform of the correlation part can be
obtained as
\begin{eqnarray}
A_{1}+A_{2}&=&\frac{8\pi^{2} \mathcal{M}G_{2}U}{(E-U-2i\epsilon
+i\gamma)}\frac{\gamma ^{2}}{(E+i\gamma -i2\epsilon) ^{2}-4\delta
^{2}}e^{(iE+2\epsilon)(x_{c}-t)}e^{(iE+2\epsilon-\gamma)\frac{\vert
x\vert }{2}}\theta(t-x_{c}). \label{FTofcorreterm}
\end{eqnarray}

According to Eqs.~(\ref{Foffreeterm}) and (\ref{FTofcorreterm}), the
second term in Eq.~(\ref{appeb1}) can be obtained by making the
replacement $x_{c}\rightarrow x_{c}$ and $x\rightarrow -x$. Then
\begin{eqnarray}
\langle x_{1},x_{2}|\psi_{rr}\rangle=-16\pi^{2}
\mathcal{M}G_{2}e^{(iE+2\epsilon)(
x_{c}-t)}\theta(t-x_{c})\phi_{rr}(x),
\end{eqnarray}
with
\begin{eqnarray}
\phi_{rr}(x)&=&t_{\delta _{1}-i\epsilon}t_{\delta
_{2}-i\epsilon}\cos( \delta x)-\frac{U}{E-U-2i\epsilon +i\gamma
}\frac{\gamma ^{2}}{(E +i\gamma-i2\epsilon) ^{2}-4\delta
^{2}}e^{\frac{(iE+2\epsilon-\gamma)}{2}\vert x\vert}.
\end{eqnarray}

Using the same method, we can obtain the wave function for the
two-photon reflection state,
\begin{eqnarray}
\langle x_{1},x_{2}|\psi_{ll}\rangle&=&\int_{0}^{\infty}
\int_{0}^{\infty}dpdqC_{p,q}^{ll}\langle
x_{1},x_{2}|\hat{l}_{p}^{\dagger }\hat{l}_{q}^{\dagger }\vert
\emptyset \rangle\approx-16\pi^{2}
\mathcal{N}G_{2}e^{-(iE+2\epsilon)(
x_{c}+t)}\theta(t+x_{c})\phi_{ll}(x),
\end{eqnarray}
with
\begin{eqnarray}
\phi_{ll}(x)&=&r_{\delta _{1}-i\epsilon}r_{\delta
_{2}-i\epsilon}\cos( \delta x)-\frac{U}{E-U-2i\epsilon +i\gamma
}\frac{\gamma ^{2}}{(E +i\gamma-i2\epsilon) ^{2}-4\delta
^{2}}e^{\frac{(iE+2\epsilon-\gamma)}{2}\vert x\vert},
\end{eqnarray}
where $\mathcal{N}$ is defined by $\langle
x_{1},x_{2}|\hat{l}_{p}^{\dagger }\hat{l}_{q}^{\dagger }\vert
\emptyset \rangle=\mathcal{N}(e^{-i\Delta_{p}x_{1}}e^{-i\Delta
_{q}x_{2}}+e^{-i\Delta _{p}x_{2}}e^{-i\Delta _{q}x_{1}})$.
\end{widetext}

\end{document}